\documentclass[useAMS,usenatbib]{mn2e}

\usepackage{graphicx}
\usepackage{longtable}
\usepackage{subfigure}
\usepackage{url}

\title[A 20~GHz bright sample for $\delta > +72^\circ$: I. Catalogue]{A~20 GHz bright sample for $\delta > +72^\circ$: I. Catalogue}
\author[S.~Righini et al.]{
\parbox[t]{\textwidth}
{S.~Righini$^{1}$\thanks{E-mail: s.righini@ira.inaf.it},
E.~Carretti$^{2}$, R.~Ricci$^{1}$, A.~Zanichelli$^{1}$, K.-H.~Mack$^{1}$, M.~Massardi$^{1}$,
I.~Prandoni$^{1}$, P.~Procopio$^{3,4}$, R.~Verma$^{1}$, M.~L{\'o}pez-Caniego$^{5}$, L.~Gregorini$^{1,6}$, F.~Mantovani$^{1}$ 
}
\vspace*{8pt} \\
$^{1}$ INAF-IRA Bologna, Via Gobetti 101, I-40129, Bologna, Italy\\
$^{2}$ CSIRO Astronomy and Space Science, PO Box 276, Parkes, NSW 2870, Australia\\
$^{3}$ INAF-IASF Bologna, Via Gobetti 101, I-40129 Bologna, Italy\\
$^{4}$ School of Physics, David Caro Building, Corner of Tin Alley \& Swanston St, University of Melbourne, Parkville, VIC 3010, Australia\\
$^{5}$ Instituto de F\'\i{sica} de Cantabria (CSIC-UC), Avda. los Castros s/n, 39005 Santander, Spain\\
$^{6}$ Dipartimento di Astronomia, Universit\'a di Bologna, via Ranzani 1, I-40127 Bologna, Italy\\
}
\begin{document}

\date{}

\pagerange{\pageref{firstpage}--\pageref{lastpage}} \pubyear{}

\maketitle

\label{firstpage}

\begin{abstract}
During 2010-2011, the Medicina 32-m dish hosted the 7-feed 18-26.5 GHz receiver built for the Sardinia Radio Telescope, with the goal to perform its commissioning. This opportunity was exploited to carry out a pilot survey at 20 GHz over the area for $\delta >~+72.3^\circ$. This paper describes all the phases of the observations, as they were performed using new hardware and software facilities. The map-making and source extraction procedures are illustrated. A customised data reduction tool was used during the follow-up phase, which produced a list of 73 confirmed sources down to a flux density of 115 mJy. The resulting catalogue, here presented, is complete above 200 mJy. Source counts are in agreement with those provided by the AT20G survey. This pilot activity paves the way to a larger project, the K-band Northern Wide Survey (KNoWS), whose final aim is to survey the whole Northern Hemisphere down to a flux limit of 50 mJy (5$\sigma$).
\end{abstract}


\section{Introduction}

Extragalactic radio sources extracted from high-frequency ($>10$~GHz) surveys are expected to have a major impact on astrophysics. They can provide samples of rare classes of sources with flat or inverted spectrum that, at low frequencies, are swamped by more numerous populations which fade away as the frequency increases (for a review see De Zotti et al. 2010). 

They can hence open a window on new classes of sources, such as those with strong synchrotron or free-free self-absorption corresponding to both very early phases of nuclear radio-activity (extreme GHz Peaked Spectrum - GPS - sources or high-frequency peakers) and late phases of the evolution of Active Galactic Nuclei (AGNs), characterized by low accretion/radiative efficiency (ADAF/ADIOS sources), as well as to early phases of the evolution of radio afterglows of gamma-ray bursts. In this context the comparison with the on-going Fermi observations will yield very interesting results (see Mahony et al. 2010).

These sources also play a vital role in the interpretation of temperature and polarisation maps of the Cosmic Microwave Background (CMB). Extragalactic point sources are one of the major foreground emissions (Planck collaboration 2011, Leach et al. 2008, Toffolatti et al. 2005); the knowledge of their positions and flux densities is crucial to remove their contribution and to estimate the residual error due to faint and unresolved components in CMB maps. As the source population composition changes at high frequency, cleaning procedures based on lower frequency catalogues are unreliable, making it essential to carry out surveys at frequencies close to the CMB window (centred at 60-70~GHz).
In addition, as the Planck satellite is now active, the realisation of a coeval 20-GHz blind survey helps, when selecting flux density limits, to avoid errors induced by high-frequency variability. 

High frequency sky surveys have become feasible very recently. Because of the faint signal, the existing surveys at 10-100~GHz usually cover small areas with good sensitivity (e.g., VSA at 34~GHz with $S_{lim}  =$ 100~mJy, Gawro\'nski et al. 2010) or consist in all-sky shallow surveys (e.g., WMAP at 23, 33, 41, 64, 94 GHz  with $S_{lim} >1$~Jy, Wright et al. 2009 a,b and ERCSC at 30, 44, 70, 100~GHz - Planck collaboration 2011 a,b,c). 
The only exception to this is the all-southern-sky Australia Telescope 20 GHz survey (AT20G), which observed the entire southern sky with the Australia Telescope Compact Array, detecting around 6000 sources down to a flux density limit of 50~mJy  (Murphy et al. 2010, Massardi et al. 2011).  

This calls for a northern sky survey with equivalent sensitivity to complete the coverage of the entire sky. Such a completeness is particularly important for several aims for which a statistical information is not sufficient, like the study of the SED of peculiar objects, the selection of samples at high radio-frequency for the northern or whole sky. A precise position of all the sources is also required to flag out the contaminated pixels from CMB maps.

The availability of a K-band (namely 18-26.5 GHz) multi-feed receiver installed on a medium-sized antenna as the Medicina 32-m dish, having a beamsize of 1.6~arcmin @~21~GHz, gave us the possibility to execute a pilot survey to verify the receiver performance, together with new software tools, while exploring a sky area which had never been extensively observed at these frequencies. This test activity paved the way to a larger project, the K-band Northern Wide Survey, which aims at performing a blind survey over the whole northern hemisphere, with a sensitivity of 50~mJy (5$\sigma$).   

The outline of the paper is the following. After a description of the system capabilities (\S \ref{sec:instrument}), we summarise in \S \ref{sec:survey} the survey strategy, including the map-making and source extraction techniques applied to achieve the list of candidate sources. The 20-GHz follow-up observations are described in \S \ref{sec:followup}, together with the automatic pipeline produced for the data reduction. Finally, the catalogue is presented in \S \ref{sec:catalogue} and results are summarised in \S \ref{sec:conclusions}.
A separate paper by Ricci et al. (hereafter 'Paper II') illustrates the detailed spectral index analysis of the sources.  

\section{The observing system}\label{sec:instrument}
The Medicina dish is a 32-m parabolic antenna located 35 km south-east of Bologna (Italy). It is operated by the Istituto di Radioastronomia (IRA), which is part of the Istituto Nazionale di Astrofisica (INAF).
The 18-26.5~GHz Multi-Feed (MF) receiver consists of seven corrugated horns in hexagonal layout, each providing Left-handed Circular Polarisation (LCP) and Right-handed Circular Polarisation (RCP) output channels, for a total of 14 channels. The instantaneous bandwidth of each channel is 2~GHz. 
Reference values for the system performances are provided in Table \ref{tab:receiver}. We notice that the decreasing gain for the lowest frequency range is due to poor illumination of the feeds: the receiver was developed for the Sardinia Radio Telescope,
consequently it was not optimised for the Medicina dish.

\begin{table}
\begin{center}
\caption{K-band Multi-Feed main features, for Elev. = $45^\circ$,  $\tau$ = 0.1.}
\begin{tabular}{cccc}
\hline
Frequency & Beamsize & $T_{sys}$ & Gain \\
GHz & arcmin & K & K/Jy \\
\hline
18 &  1.7 &  43 & 0.10 \\
22 &  1.4 &  72 & 0.11 \\
26 &  1.2 &  79 & 0.11 \\ 
\hline
\label{tab:receiver}
\end{tabular}
\end{center}
\end{table}
 
The broad total bandwidth delivered by the MF outputs (28~GHz) is detected by an analogue total power backend, expressly produced for its use with the MF receiver.  
The main specifications of this backend are summarised in Table \ref{tab:backend}.
Both these hardware components were developed by the Institute of Radioastronomy (INAF) for the new Sardinia Radio Telescope, but were installed on the Medicina dish to undergo the commissioning phase.

\begin{table}
\begin{center}
\caption{Total power backend characteristics.}
\begin{tabular}{ll}
\hline
IF inputs & 14 x 3, in the range 0.1-2.1 GHz \\ 
IF outputs & 2, in the range 0.1-2.1 GHz \\
Instant bandwidth & Selectable: 150, 680, 1200, 2000 MHz \\
Cable equalisation & Up to 12 dB \\
Attenuators & Variable: 0-15 dB \\
Resolution & Up to 21 bit \\
Sampling interval & 0.001-1 s \\
Noise source & Chopping frequency: 0.5 - 500 Hz \\
\hline
\label{tab:backend}
\end{tabular}
\end{center}
\end{table}


\section{Survey test observations}\label{sec:survey}

In winter 2010 we produced a total intensity test map covering the Northern polar cap ($\sim 880$ deg$^2$, $\delta > 72.3^\circ$ ), which allowed us to deeply check the new hardware and software facilities - the MF, the analogue continuum backend and the telescope control system - which is a new system based on the Alma Common Software (see Orlati et al. 2012). 
The actually chosen band, due to RFI constraints, was 20-22~GHz. 

\begin{figure}
\centering
\resizebox{\hsize}{!}{\includegraphics{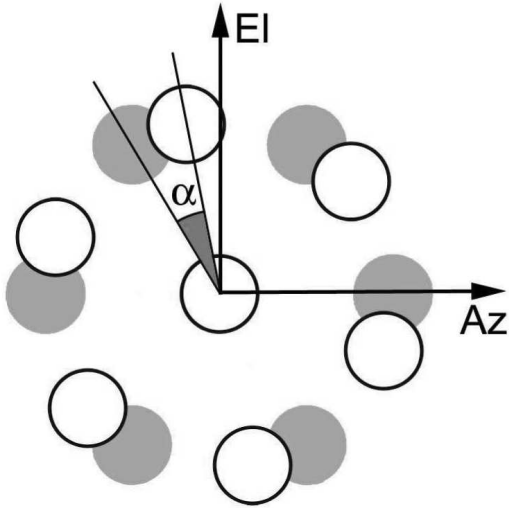}}
\caption{The projected beams in the Az-El frame. Empty circles correspond to the rotated position, filled circles to the rest position.}
\label{fig:feeds}
\end{figure}

The observing strategy consisted in long and fast ($15^\circ$/min) On-The-Fly (OTF) azimuth scans - i.e. at constant elevation. 
This technique was based on that developed for the project S-PASS at the Parkes radiotelescope (Carretti 2010, Carretti et al. in prep.). 
To optimise the area scanned by the seven beams, the MF array was rotated by an angle $\alpha=19.1^\circ$ with respect to its rest position. This way, the paths run by the individual beams were equally spaced in elevation realising a regular sampling of the sky.
Figure \ref{fig:feeds} shows the projected beams in the Horizontal reference frame.

The apparent rotation of the celestial sphere was exploited to cover the sky area to map, which was a 'Declination strip' spanning 24 hours in Right Ascension. In a single sidereal day this technique allowed us to observe, in the Equatorial frame, 'wavy' stripes within the belt - see figure 8 in Carretti (2010). 
To complete the map this scheme was repeated in the following days, shifting the stripe pattern to fill in the missing areas until the Nyquist sampling, at least, was reached. In practice, this translated into starting the azimuth scans sequence at a different LST. The sidereal-time interleave between two adjacent sequences was computed so that the stripes covered by the scans were spaced by half a beamsize or less. In particular, the interleave produced the Nyquist sampling for the outer region of the map, while the inner area reached a higher sampling.

Observing at a constant elevation is extremely helpful for both data stability and calibration. Ground emission is mainly elevation-dependant, and the same holds for atmospheric opacity: when observing in steady, uniform weather conditions such as in a dry, clear day, the opacity is only a function of the airmass. \newline The flux density calibration of the map was achieved observing calibrators (3C286, 3C295, 3C48, 3C123) at least four times a day, when each of them transited at the same elevation of the horizontal scans. As weather conditions during the observations were quite stable, this simplification was acceptable to calibrate each of the 24-hour-long datasets. 

Fast scans allowed us to cover the chosen polar cap area with declination in the range between $[+72.3^\circ,+89^\circ]$ in 4 days.  Specifically, scans were performed at $15^ \circ$/min between $1^\circ$ and $25^\circ$ of azimuth, taken at elevation $44.52^\circ$.

\subsection{Theoretical sensitivity} 
A single On-The-Fly (OTF) scan has a theoretical instantaneous noise which can be computed by means of the radiometer equation (as in Rohlfs \& Wilson):

\begin{equation}
\sigma_S= \frac{kT_{sys}}{G\sqrt{B t}} [Jy/beam]
\label{eq:radiometer}
\end{equation}

where \textit{k}=1 for a single polarisation, \textit{G} is the antenna gain (K/Jy), \textit{B} is the bandwidth (Hz), \textit{t} is the integration time (s).  
As a reference value to assess the expected performance of the system we use the one-second integration sensitivity, which reads $\sigma_{1s} = 14.6$~mJy/beam for a single polarisation, for observations carried out at 21~GHz with a bandwidth of 2~GHz (values for $T_{sys}$ and gain are extrapolated from Table \ref{tab:receiver}).  

We now achieve a first order estimate of the theoretical noise level of a map obtained exploiting the above mentioned scanning strategy. The following computation refers to the integration of OTF scans in the 20-22~GHz band, performed at the constant elevation \textit{El}$=45^\circ$, considering that:
\begin{itemize}
\item the actual scan speed on sky is $15^\circ$/min$\cdot$cos(\textit{El}) = 10.7$^\circ$/min;
\item one beam-sized pixel (1.6~arcmin) is observed for 0.150~s in each individual subscan; 
\item the scanning strategy includes back and forth scans. Along with the proper spacing between adjacent subscans (half a beamwidth), this leads to observe a beam-sized pixels for four times, for a total integration time $t=0.600$~s; 
\item every feed observes two circular polarisations, which further improves the sensitivity by a factor of $\sqrt{2}$, since $k=\frac{1}{\sqrt{2}}$ for dual polarisation.
\end{itemize}

Equation \ref{eq:radiometer} thus yields $\sigma_S =13.4$~mJy/beam, as concerns the less sampled area of the map - i.e. its southern edge, see \ref{sec:map}.

\subsection{Map-making}\label{sec:map}

We built the survey map with a custom software, adapted from that of S-PASS for compact source detection (see Carretti 2010). 
For every OTF scan, data streams relative to the various feeds were detached and 
individually calibrated. To optimise for compact source detection, 
a high-pass median filtering was applied to each data stream 
to subtract its baseline (as suggested in Gregory et al. 1996). In particular, the filter width was set to 6.4 arcmin (corresponding to 0.6 seconds along the scans). 
This effectively removed the large scale
sky signal component, together with most of the ground emission and atmospheric fluctuations.

Data were binned together in two different ways to be
compliant with different data analysis tools. A first map
was realised binning data in a HEALPix pixelation
(G{\'o}rski et al. 2005). The angular resolution parameter {\it nside}
was set to 4096 for pixels of size of 0.85'. A second map was
realised binning data on a grid of 0.8'x0.8'. The gridding was
performed in a Zenithal equidistant projection
(ARC, Calabretta \& Greisen 2002) centred at
the North Celestial Pole.

\begin{figure}
 \centering
 \resizebox{\hsize}{!}{\includegraphics[angle=0]{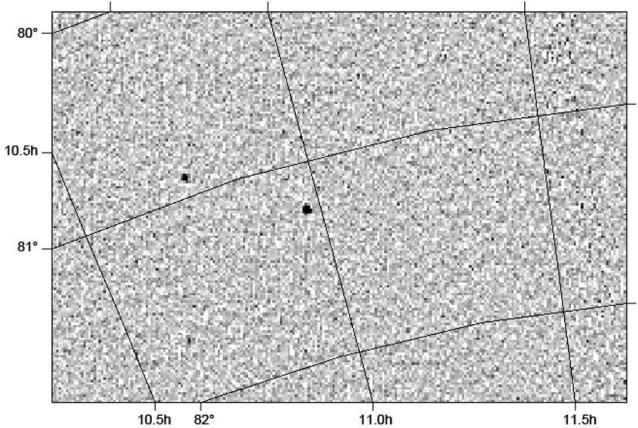}}
 \caption{Small patch of the map, showing two bright sources.}
\label{fig:map}
\end{figure}

During the observations, one of the seven feeds was excluded due to hardware instabilities. 
This, together with the non-ideal atmospheric opacity and the occasional presence of RFI, led to a slightly higher-than-predicted 
noise in the final map. 
The scanning geometry produced an inhomogeneous sampling and a different integration time, both Declination-dependant, over the observed area. 
In particular, the high-declination inner core of the map 
reached a noise level of 10~mJy/beam, while the map southern edge was limited to 20~mJy/beam. 
The median noise level was 15.4~mJy/beam.
A small patch of the map with two obvious bright sources is illustrated in Figure \ref{fig:map}.
The scan interleave had been, due to time constraints, set to the minimum Nyquist sampling (half the
beamsize). This was sufficient for source detection but not suitable to obtain a photometric map for compact sources. As a consequence, 
follow up observations were required for an accurate estimate of the flux densities.

\subsection{Source extraction techniques}\label{sec:srcextraction}
Three source extraction tools were employed to try the detection of candidate sources on our map: 
\begin{itemize}
\item ifcaMex;
\item Signal-to-Noise-Ratio Source Extraction (SNRSE);
\item SExtractor.
\end{itemize}

The first software, ifcaMex, was provided by Marcos L{\'o}pez-Caniego. This code was developed in the light of the source extraction needs for maps obtained with the Planck satellite observations. It had demonstrated its reliability on WMAP 5-years maps (Massardi et al. 2009), but it had never been applied to ground-based single dish data. Details on ifcaMex can be found in Gonz{\'a}lez-Nuevo et al. (2006). \newline
As concerns SNRSE, custom produced, here follows a brief description of its features. In order to extract the candidate sources from the surface brightness 
map, a sensitivity map was obtained by computing the rms noise in a 20-pixel-wide box centred around each pixel of the surface brightness map. By matching the sensitivity map with the surface brightness map, a list of bright pixels (with SNR $>5$) was extracted. Neighbouring bright pixels were removed within a radius of one HPBW starting from the brightest and moving towards the faintest ones
in order to single out candidate source positions. Finally, to improve positional and flux density accuracy, a 2D Gaussian fit was performed for each candidate source. \newline 
SExtractor (Bertin et al. 1996) was employed as a third tool to extract the candidate sources positions. \newline
The three methods underwent a comparison test over semi-synthetical maps. We first extracted from our map a raw list of detections with each method. We built a background map, removing from the actual map all the candidate sources and replacing them with noise pixels copied from the surroundings. Synthesized point-like sources were then injected. These sources were produced following the expected counts and flux density distribution. In total, we injected 228 sources ranging from 50~mJy to 5~Jy. We generated 10 different maps, randomly changing the spatial distribution of the sources, and performed the extraction on each of them with the three codes.
Table \ref{tab:extraction} summarises the results. 

\begin{table}
\begin{center}
\caption{Average efficiency of the extraction methods on the test maps, computed for various flux density limits.}
\begin{tabular}{ccccc}
\hline
$S_{lim}$ & Injected & ifcaMex & SNRSE & SExtractor \\
Jy & sources & detections & detections & detections \\
\hline
0.05 & 228 &   19.8   &   35.7   &   75.4 \\
0.10 & 126 &   34.5   &   60.1   &   97.5 \\
0.20 &  56  &   67.9   &   95.2   &   99.8 \\
0.50 &  16  &   97.1   &   99.4   & 100.0 \\
1.00 &   3   & 100.0   & 100.0   & 100.0 \\ 
1.50 &   3   & 100.0   & 100.0   & 100.0 \\ 
\hline
\label{tab:extraction}
\end{tabular}
\end{center}
\end{table}

These tests indicated SExtractor to perform better than the other methods. ifcaMex missed most of the sources with $S <200$ mJy; further tests are desirable to investigate the software performance when varying its setup parameters, in order to better match the features of a map built like ours. It must be noticed that, due to the distinct filters employed, each code actually listed a set of candidate sources different from the others. All the methods produced a large raw list of candidates ($>4000$ for SExtractor and ifcaMex, only about 2000 for SNRSE), mainly located in the outer regions of the map, where the increasing noise adds to the effects caused by the low resolution of the map. \\
We then compared the detection lists achieved on the original map, containing the real sources.  Discrepancies were already present at flux densities $>150$~mJy (10$\sigma$). Visually checking the detections on the map, we verified that none of the single methods alone was able to detect all the evident sources. This prevented us from being able to select a unique tool for the extraction. \\ 
For the sake of the follow-up phase, the three detection lists were merged and cleaned on the basis of an accurate visual inspection of the map. We easily removed artefacts due, for example, to short-time instabilities affecting the feeds, which turned into very clear marks on the map. Other features were present in areas affected by RFI.  We also rejected candidates having a profile incompatible with the beamsize. Due to commissioning constraints limiting the available time for the follow-up phase, the list was further restricted, my means of a stricter visual inspection, to a final selection of 151 candidate sources. This process likely introduced a bias against the faintest sources. 

\section{20 GHz follow-up}\label{sec:followup}

Follow-up observation were carried out between December 2010 and February 2011. They were performed in the 19-21~GHz band, to cope with RFI.
Sources were almost contemporarily observed at 5~GHz. Candidates confirmed at either one of the two frequencies were then observed at 8~GHz; these multi-frequency data are discussed in Paper II. 
Further high-frequency observations were performed, for all the sources which had been previously confirmed, in April 2011. Candidates were this time observed using a reduced bandwidth (19.50-20.18~GHz, HPBW=1.7~arcmin) in order to further mitigate the RFI-related effects. The source catalogue presented in this paper lists the flux densities obtained in this final session.

\begin{table}
\begin{center}
\caption{Follow-up scan setup.}
\begin{tabular}{cccc}
\hline
Frequency &  Length & Speed & Sampling \\
GHz & arcmin & arcmin/s & s \\
\hline
19.50-20.18 & 12.5 & 1.0 & 0.040 \\ 
\hline
\label{tab:otf}
\end{tabular}
\end{center}
\end{table}

\begin{figure}
\centering
\resizebox{\hsize}{!}{\includegraphics[angle=90]{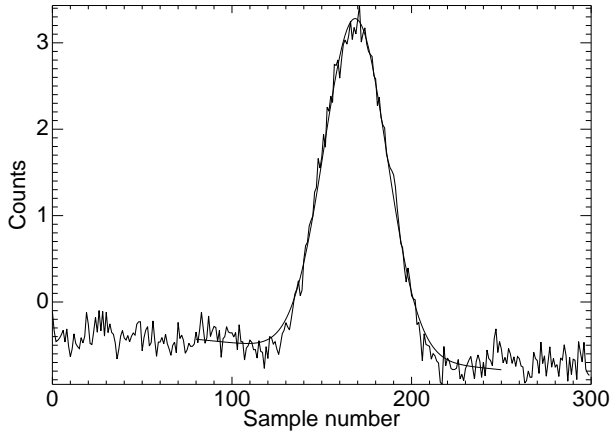}}
\caption{Portion of a single, raw OTF scan over the brightest source of the catalogue.}
\label{fig:source1}
\end{figure}

\begin{figure}
\centering
\resizebox{\hsize}{!}{\includegraphics[angle=90]{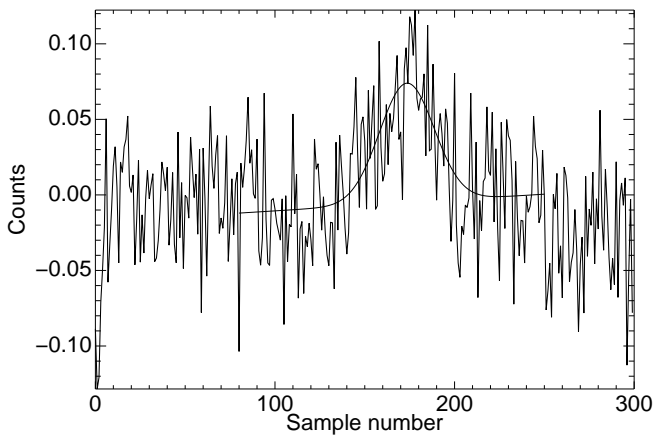}}
\caption{Portion of a single, raw OTF scan over the faintest source of the catalogue.}
\label{fig:source335}
\end{figure}

Table \ref{tab:otf} summarises the scan parameters chosen for follow-up observations.
Each scan, an On-The-Fly acquisition along Right Ascension or Declination across the source, produced an output FITS file which contained, together with several tables listing the system setup parameters, two raw data streams (or channels) in arbitrary counts. They corresponded to the LCP and the RCP total intensities, in this case coming from the central feed of the MF receiver.  
All the acquired scans were visually inspected and flagged, in order to select only the suitable scans/channels for the subsequent data reduction phase. 
Given the point-like nature of the sources, they appeared in the scans like Gaussian profiles reproducing the antenna pattern main beam, embedded in a baseline having a linear profile (see Figures \ref{fig:source1} and \ref{fig:source335}). Data flagging focused, as a consequence, on the selection of those scans which showed little or no alteration of the baseline and of the Gaussian profile, in order to allow a smooth integration and a good fit during the data reduction. 

\subsection{Data reduction}\label{sec:datared}

Data reduction was performed using an updated version of the OTF Scan Calibration-Reduction (OSCaR) pipeline, a customisable ensemble of IDL routines capable of handling large datasets, operating at all the frequencies available using the Medicina 32-m dish (see Procopio et al. 2011 for details).
The main steps performed by this release of OSCaR can be summarised as follows: 

\begin{enumerate}
\item first estimation of the factor to directly convert the raw signal, which is stored in arbitrary counts, to flux density units. This is obtained computing the ratio between the catalogue flux density of a calibrator (Jy) and the raw amplitude measured by means of a Gaussian fit on the scans performed on the calibrator (counts). These conversion factors are then extrapolated to the reference elevation of $90^\circ$, exploiting the antenna gain curve. They are also rescaled for the atmospheric absorption thanks to the opacity values, which are measured from actual skydip scans obtained right after the calibration scans. This way, the conversion factors can be considered as normalised; 
\item reconstruction of a timeline to record the variation of the normalised counts-to-Jy factor and estimate the calibration-related error on flux density;
\item association of an average counts-to-Jy factor to each set of contiguous scans on a candidate source, on the basis of the observation time. The value is then rescaled applying the antenna gain curve, i.e. taking into account the actual elevation at which the source had been observed, and the atmospheric opacity variations, which might have taken place between the calibration observation and the source observation. This conversion factor is applied once the contiguous scans have been integrated;  
\item fitting of a Gaussian profile over the integrated scan to measure the source flux density and estimate its uncertainty. If the source had been observed in more than one session within the analysed dataset, also the globally integrated flux density is reported;
\item rescaling of the measured flux density, taking into account the positional offset between the position extracted from the map - and used to performed the cross-scans - and the one determined by fitting the cross-scans. This offset was generally very small, but in some cases it was as large as half the beamsize - since this was the pixel size of the map. 
\end{enumerate} 

As both the initial data flagging and the pipeline internal checks were performed on the single channels (LCP and RCP, for each scan), raw data for every cross-scan was composed by four independent estimates: LCP and RCP taken from either Right Ascension and Declination scans.  
OSCaR provides an output table which lists, for each source, all the intermediate results: from the four separate estimates to the partial integrations, performed by channel and by scan direction, to the final flux density measurement. Each integration step consists in averaging the scans, weighing over their uncertainties. These uncertainties include three different contributions: 1) error due to calibration; 2) error on the amplitude measured by the Gaussian fit (usually negligible); 3) error due to the pointing offset. The major contribution is the first (about $5\%$ on average). 
Thanks to the possibility of inspecting all the internal phases of the process, the final flux density measurements underwent a further selection phase. In particular, we rejected the measurements showing: 
\begin{itemize}
\item scans available in one direction only (RA or Dec);
\item scans available in one channel only (LCP and RCP);
\item SNR $< 5$;
\item an unrealistic ratio between the LCP and RCP flux densities, indicating RFI contamination.
\end{itemize}

\begin{figure}
 \centering
 \resizebox{\hsize}{!}{\includegraphics[angle=90]{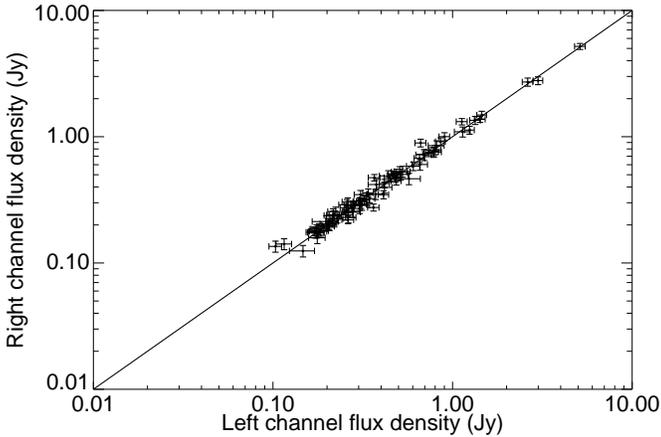}}
 \caption{RCP vs LCP flux densities for the single measurements.}
 \label{fig:RvsL}
\end{figure}

The final list contained 73 confirmed sources out of 151 candidates.
Most of the sources had been observed at least twice. At the end of the selection, 21 sources preserved multiple measurements, which were averaged to obtain a final flux density for the catalogue. Figure \ref{fig:RvsL} shows the LCP and RCP flux densities for the single measurements, prior to the final averaging.

Absolute calibration was achieved considering, for the calibrators nominal flux densities, the values obtained with the polymonial models given by Baars et al. (1977). The reliability of these models was confirmed by recent observations carried out at Effelsberg (Kraus, 2010, private communication). 
The calibration procedure within OSCaR required at least two observations over a calibrator in one 24-hour session. This goal was largely met, with the only exception of April, 15th, when a unique session over a calibrator positively passed the flagging phase. For that specific day, a flat calibration error equal to 5$\%$ was assigned to all the measurements.    

\section{20-GHz catalogue }\label{sec:catalogue}

\begin{figure}
 \centering
 \resizebox{\hsize}{!}{\includegraphics[angle=90]{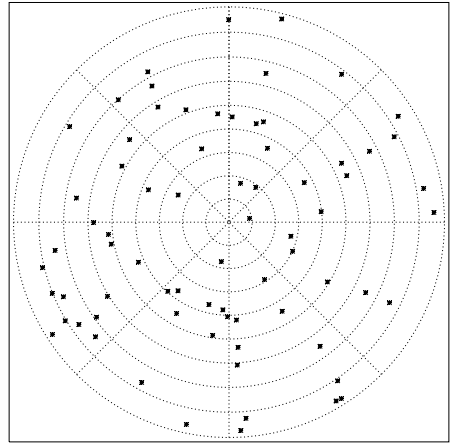}}
 \caption{Positions of the confirmed sources. Map is centred on the North Celestial Pole, with the outer rim being Dec=72$^\circ$. The RA=0h meridian is the radius pointing to the bottom side of the plot.}
\label{fig:display}
\end{figure}

\begin{figure}
 \centering
 \resizebox{\hsize}{!}{\includegraphics[angle=90]{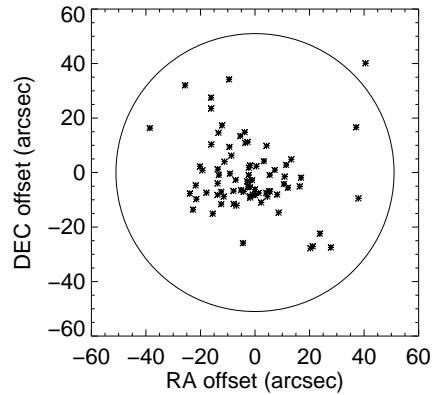}}
 \caption{Position error with respect to the NVSS counterparts. The circle is one beamsize in diametre (102~arcsec).}
\label{fig:posmat}
\end{figure}

The flux densities of the 73 confirmed sources range from 115~mJy to 5~Jy. Figure \ref{fig:display} shows where the sources are located.
The positional accuracy
was checked using the NVSS catalogue (Condon et al. 1998). A cross-match 
by position between the two catalogues within a search radius of 102~arcsec (one beam size)
provided full matches for our catalogue (all sources have
a NVSS counterpart).
The displacements between the positions of our sources 
and their NVSS counterparts are shown in Figure \ref{fig:posmat}. The rms displacements in RA
and Declination are 15 arcsec and 13 arcsec respectively, not far from the expected positional error of $\simeq 1/10$ of the beamsize.

\begin{figure}
\centering
\resizebox{\hsize}{!}{\includegraphics[angle=90]{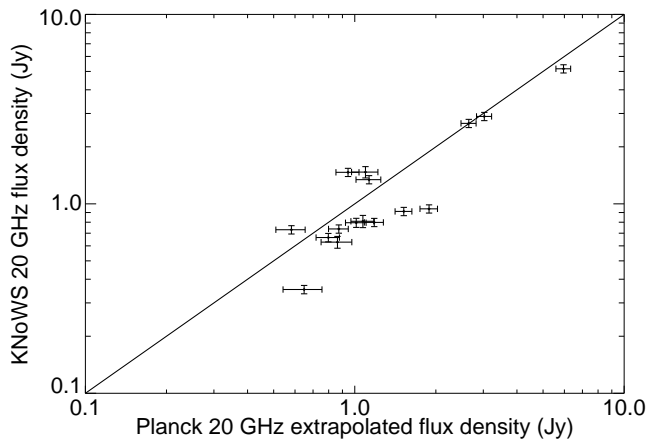}}
\caption{Comparison between KNoWS flux densities and Planck flux densities extrapolated to 20~GHz. Individual spectral indices were applied to each source (median value is -0.39).}
\label{fig:kvsp}
\end{figure}

This bright catalogue was cross-matched with the 30~GHz data from the Planck ERCSC catalogue, obtaining 16 matches. Planck 30-GHz flux densities were then rescaled to 20~GHz according to individual spectral indices. These, in turn, were computed exploiting 30-GHz follow-up observations performed (in autumn 2010) over our sources using the OCRA system at the Torun radio telescope - presented in Paper II together with details on the statistical and spectral properties of the sources. The resulting flux density comparison is illustrated in Figure \ref{fig:kvsp}. The linear best fit is $y = 1.058x + 0.100$ (uncertainties on the two parameters are 0.084 and 0.154 respectively). Taking into account that none of the three measurements (Medicina, Planck, Torun) were strictly coeval, the observed dispersion can be explained by the variability of the sources, which has a typical time-scale of weeks to months. 
As a check, we were able to compare the flux densities of sources KNOWS021733+734923, KNOWS041050+765649 and KNOWS180044+782812 (see Table \ref{tab:catalogue}) with literature data. The flux densities agree within a few percents.

\subsection{Source counts}\label{sec:srccnt}

\begin{figure}
\centering
\resizebox{\hsize}{!}{\includegraphics[angle=90]{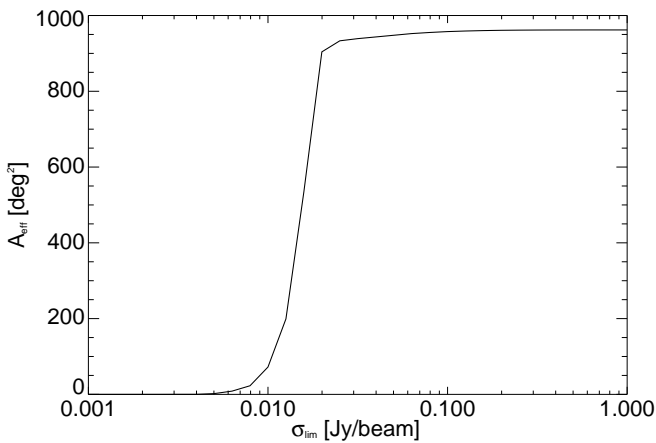}}
\caption{Effective area as a function of the map noise level.}
\label{fig:aeff}
\end{figure}

Source counts at 20~GHz were determined using the flux densities of the
73 sources confirmed by follow-up observations. The 
sensitivity map, previously used to extract candidate sources with SNRSE, was here employed
to compute the visibility function of the survey - i.e. the effective area 
covered as a function of the survey flux density ($A_{\rm eff}(>S)$). Figure \ref{fig:aeff} shows how $A_{\rm eff}$ varies with the map noise level. 

\begin{figure}
 \centering
 \resizebox{\hsize}{!}{\includegraphics[angle=90]{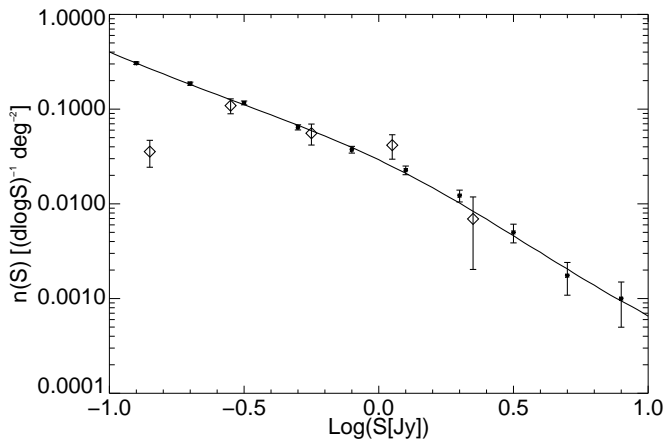}}
 \caption{Differential source counts. Empty diamonds correspond to our catalogue, asterisks to
 the AT20G data. The solid line shows the counts predicted with the model by De Zotti et al.}
 \label{fig:counts} 
\end{figure}

The counts were logarithmically binned in flux density starting from the 
faintest source in the catalogue. The differential counts $n_i$ as a function
of flux density have been derived as follows:

\begin{equation}
n_i = \frac{1}{\Delta log(S)}\ \sum_{j}^{N_i} \frac{1}{A_{eff}(S_j)}
\label{eq:robeqa}
\end{equation}

\begin{equation}
log(S_i) \leq log(S_j) < log(S_i) + \Delta log(S)
\label{eq:robeqb}
\end{equation}

\begin{equation}
\Delta n_i = \frac{\sqrt{N_i}}{A_{eff(S_i)}\ \Delta log(S)}
\label{eq:robeqc}
\end{equation}

where $n_i$ in Equation \ref{eq:robeqa} represents the number of sources $N_i$ in the logarithmic
flux density bin $i$ defined in Equation \ref{eq:robeqb}. The counts were weighted for the 
effective area $A_{\rm eff}(S_j)$ visible by each source $j$. $\Delta n_i$ is
the Poissonian error to the counts weighted for the A$_{\rm eff}$ at the 
flux density $S_i$ of the bin centre. The differential counts of our catalogue
 in Figure \ref{fig:counts} are compared with the model counts by de Zotti et al. 
(2005) and the counts from the AT20G Full Sample catalogue. 
A good agreement can be seen between our counts 
and both the model and the AT20G counts down to the second to faintest flux
density bin. Table \ref{tab:Intcounts} instead lists integral counts, again weighted for the 
effective area: our counts are compared to the ones obtained from De Zotti's model and the counts resulting within the AT20G southern polar cap (the homologous area to our survey, i.e. $-89^\circ < \delta < -73.2^\circ$), confirming that the completeness of our catalogue is limited to 200 mJy. This is most likely due to the bias introduced by the selection of the sources (see \ref{sec:srcextraction}).

\begin{table}
\begin{center}
\caption{Comparison among theoretical and observed integral counts, weighted for the effective area. Poissonian errors are given in brackets for the observed counts.}
\begin{tabular}{cccc}
\hline
$S_{lim}$ &  De Zotti & KNoWS & AT20G \\
mJy &  &  &  \\
\hline
100 & 145 & 73 (9)& 125 (11)   \\ 
200 &   63 & 63 (8)& 57 (8)   \\
\hline
\label{tab:Intcounts}
\end{tabular}
\end{center}
\end{table}

\section{Conclusions}\label{sec:conclusions}
Our project served as a fundamental test for the hardware and software facilities which were being commissioned in Medicina. The survey/mapping initial phase and the subsequent multi-frequency follow-up allowed us to deeply test the main continuum observing modes. They also helped us in improving our knowledge of the influence of RFI over high-frequency acquisitions performed with unprecedented sensitivity, as these single-dish facilities were not previously available in Medicina. 
Map-making, source extraction and data reduction tools, both known and custom-developed, were tested and debugged, allowing us to fine tune the whole processing phase going from data acquisition to the reduction and calibration of high-frequency maps and cross-scans (including the crucial issues related to atmospheric opacity). 
The survey, performed exploiting an ad hoc observing technique, produced a shallow map of the region for $\delta > 72.3 ^\circ$, with an average noise level of 15.4~mJy. 
A selection of the extracted source candidates was followed-up, leading to the confirmation of 73 sources, down to a flux density of 115~mJy. The characteristics of the map, together with a strict selection of the source candidates imposed by the short commissioning time available, produced this final catalogue of bright sources, which is complete for $S_{lim} = 200$~mJy. It must be noticed that this region of the sky had never been extensively observed at 20~GHz down to this flux density limit, thus this catalogue constitutes a useful reference for spectral studies of the listed sources. A separate paper (Ricci et al., in prep.) illustrates the multi-frequency observations and spectral analysis carried out for these sources.

\section*{Acknowledgments}
This work is based on observations performed with the Medicina telescope, operated by INAF - Istituto di Radioastronomia. We gratefully thank the staff at the Medicina radio telescope for the valuable support they provided. A particular acknowledgment goes to Andrea Orlati, Andrea Maccaferri and Alessandro Orfei.
We warmly thank Uwe Bach and Alex Kraus of the 100-m Effelsberg telescope of the Max Planck - Institut f\"{u}r Radioastronomie who provided us with important feedback on the flux densities of calibrators prior to publication.

\bsp

\newpage
\begin{table*}
\begin{center}
\caption{The KNoWS pilot bright sample. Continues on next page.}
\label{tab:catalogue}
\begin{tabular}{lrrrrr}
\hline
NAME & ID &  RA & DEC & $S_{20GHz}$ & $\sigma_{20GHz}$ \\
& & [hh:mm:ss]&[dd:mm:ss.s]& mJy & mJy\\
\hline
KNOWS001302+723123 &  1 &  00:13:02 & 72:31:22.8 &     325 &      18 \\
KNOWS001312+774854 &  2 &  00:13:12 & 77:48:54.0 &     208 &      19 \\
KNOWS001633+791641 &  3 &  00:16:33 & 79:16:40.8 &     247 &      12 \\
KNOWS001716+813441 &  4 &  00:17:16 & 81:34:40.8 &     745 &      38 \\
KNOWS001948+732725 &  5 &  00:19:48 & 73:27:25.2 &    1283 &      64 \\
KNOWS020313+810622 &  6 &  02:03:13 & 81:06:21.6 &     174 &       9 \\
KNOWS020336+723253 &  7 &  02:03:36 & 72:32:52.8 &     594 &      34 \\
KNOWS020649+841102 &  8 &  02:06:49 & 84:11:02.4 &     120 &      12 \\
KNOWS020954+722920 &  9 &  02:09:54 & 72:29:20.4 &     488 &      42 \\
KNOWS021733+734923 & 10 &  02:17:33 & 73:49:22.8 &    2894 &     145 \\
KNOWS022459+765544 & 11 &  02:24:59 & 76:55:44.4 &     130 &      11 \\
KNOWS035447+800918 & 12 &  03:54:47 & 80:09:18.0 &     294 &      21 \\
KNOWS041050+765649 & 13 &  04:10:50 & 76:56:49.2 &    1342 &      67 \\
KNOWS041318+745107 & 14 &  04:13:18 & 74:51:07.2 &     297 &      25 \\
KNOWS042132+835837 & 15 &  04:21:32 & 83:58:37.2 &     205 &      10 \\
KNOWS050844+843202 & 16 &  05:08:44 & 84:32:02.4 &     301 &      15 \\
KNOWS061048+724843 & 17 &  06:10:48 & 72:48:43.2 &     360 &      25 \\
KNOWS062555+820228 & 18 &  06:25:55 & 82:02:27.6 &     628 &      45 \\
KNOWS063921+732454 & 19 &  06:39:21 & 73:24:54.0 &    1467 &      73 \\
KNOWS064132+881200 & 20 &  06:41:32 & 88:12:00.0 &     204 &      14 \\
KNOWS072608+791135 & 21 &  07:26:08 & 79:11:34.8 &     498 &      25 \\
KNOWS074713+763918 & 22 &  07:47:13 & 76:39:18.0 &     498 &      25 \\
KNOWS074922+742038 & 23 &  07:49:22 & 74:20:38.4 &     416 &      21 \\
KNOWS075039+790914 & 24 &  07:50:39 & 79:09:14.4 &     240 &      12 \\
KNOWS075052+824200 & 25 &  07:50:52 & 82:42:00.0 &     475 &      24 \\
KNOWS080817+731514 & 26 &  08:08:17 & 73:15:14.4 &     320 &      16 \\
KNOWS092934+861236 & 27 &  09:29:34 & 86:12:36.0 &     187 &       9 \\
KNOWS093056+742017 & 28 &  09:30:56 & 74:20:16.8 &     240 &      15 \\
KNOWS101009+825020 & 29 &  10:10:09 & 82:50:20.4 &     353 &      18 \\
KNOWS104421+805447 & 30 &  10:44:21 & 80:54:46.8 &    1188 &      59 \\
KNOWS105359+863004 & 31 &  10:53:59 & 86:30:03.6 &     126 &       9 \\
KNOWS105812+811438 & 32 &  10:58:12 & 81:14:38.4 &     795 &      44 \\
KNOWS110149+722544 & 33 &  11:01:49 & 72:25:44.4 &     912 &      46 \\
KNOWS110410+765859 & 34 &  11:04:10 & 76:58:58.8 &     278 &      14 \\
KNOWS115311+805837 & 35 &  11:53:11 & 80:58:37.2 &     880 &      44 \\
KNOWS120019+730054 & 36 &  12:00:19 & 73:00:54.0 &     750 &      38 \\
KNOWS122340+804016 & 37 &  12:23:40 & 80:40:15.6 &     479 &      24 \\
KNOWS132143+831623 & 38 &  13:21:43 & 83:16:22.8 &     395 &      28 \\
KNOWS132351+794258 & 39 &  13:23:51 & 79:42:57.6 &     378 &      19 \\
KNOWS135324+753307 & 40 &  13:53:24 & 75:33:07.2 &     341 &      17 \\
KNOWS135756+764330 & 41 &  13:57:56 & 76:43:30.0 &     458 &      23 \\
KNOWS140638+782816 & 42 &  14:06:38 & 78:28:15.6 &     115 &       6 \\
KNOWS144830+760137 & 43 &  14:48:30 & 76:01:37.2 &    1225 &      61 \\
KNOWS152107+785837 & 44 &  15:21:07 & 78:58:37.2 &     178 &       9 \\
KNOWS155608+742107 & 45 &  15:56:08 & 74:21:07.2 &     190 &      11 \\
KNOWS160731+850159 & 46 &  16:07:31 & 85:01:58.8 &     265 &      13 \\
KNOWS160922+794023 & 47 &  16:09:22 & 79:40:22.8 &     287 &      22 \\
KNOWS163235+823228 & 48 &  16:32:35 & 82:32:27.6 &     798 &      40 \\
KNOWS172404+765328 & 49 &  17:24:04 & 76:53:27.6 &     730 &      37 \\
KNOWS180044+782812 & 50 &  18:00:44 & 78:28:12.0 &    2660 &     133 \\
KNOWS182316+793856 & 51 &  18:23:16 & 79:38:56.4 &     279 &      14 \\
KNOWS183659+750741 & 52 &  18:36:59 & 75:07:40.8 &     257 &      14 \\
KNOWS184218+794540 & 53 &  18:42:18 & 79:45:39.6 &     941 &      47 \\
KNOWS185458+735129 & 54 &  18:54:58 & 73:51:28.8 &     263 &      14 \\
KNOWS192754+735816 & 55 &  19:27:54 & 73:58:15.6 &    5165 &     258 \\
KNOWS193526+813022 & 56 &  19:35:26 & 81:30:21.6 &     193 &      10 \\
KNOWS193706+744102 & 57 &  19:37:06 & 74:41:02.4 &     420 &      21 \\
KNOWS200422+735505 & 58 &  20:04:22 & 73:55:04.8 &     308 &      15 \\
KNOWS200539+775252 & 59 &  20:05:39 & 77:52:51.6 &     809 &      60 \\
KNOWS200955+722920 & 60 &  20:09:55 & 72:29:20.4 &     737 &      37 \\
\hline
\end{tabular}
\end{center}
\end{table*}

\newpage
\setcounter{table}{2}
\begin{table*}
\begin{center}
\label{tab:cataloguecont}
\begin{tabular}{lrrrrr}
\hline
NAME & ID &  RA& DEC & $S_{20GHz}$ & $\sigma_{20GHz}$ \\
& & [hh:mm:ss]&[dd:mm:ss.s]& mJy & mJy\\
\hline
KNOWS201716+744059 & 61 &  20:17:16 & 74:40:58.8 &     464 &      34 \\
KNOWS202242+761131 & 62 &  20:22:42 & 76:11:31.2 &    1471 &     101 \\
KNOWS204240+750802 & 63 &  20:42:40 & 75:08:02.4 &     229 &      12 \\
KNOWS211402+820437 & 64 &  21:14:02 & 82:04:37.2 &     235 &      12 \\
KNOWS213345+823904 & 65 &  21:33:45 & 82:39:03.6 &     302 &      15 \\
KNOWS220039+805844 & 66 &  22:00:39 & 80:58:44.4 &     166 &      12 \\
KNOWS220549+743632 & 67 &  22:05:49 & 74:36:32.4 &     201 &      10 \\
KNOWS230524+824232 & 68 &  23:05:24 & 82:42:32.4 &     243 &      17 \\
KNOWS231226+724055 & 69 &  23:12:26 & 72:40:55.2 &     310 &      23 \\
KNOWS231556+863130 & 70 &  23:15:56 & 86:31:30.0 &     280 &      19 \\
KNOWS232713+801236 & 71 &  23:27:13 & 80:12:36.0 &     213 &      16 \\
KNOWS234405+822638 & 72 &  23:44:05 & 82:26:38.4 &     309 &      15 \\
KNOWS235626+815255 & 73 &  23:56:26 & 81:52:55.2 &     664 &      34 \\
\hline
\end{tabular}
\end{center}
\end{table*}

\end{document}